\documentstyle[aps,floats,twocolumn,prd,psfig]{revtex}

\newcommand{\ApJ}{Astrophys. J.}

\newcommand{\MNRAS}{Mon. Not. Roy. Astron. Soc.}

\def\sun{\hbox{$\odot$}}

\newcommand{\bn}{\hat{\bf n}}

\newcommand{\vecl}{{\bf l}}

\newlength{\tskip}
\newlength{\colwidth}

\newcommand{\beq}{\begin{equation}}
\newcommand{\eeq}{\end{equation}}
\newcommand{\beqa}{\begin{eqnarray}}
\newcommand{\eeqa}{\end{eqnarray}}
\def\VEV#1{{\langle #1 \rangle}}
\long\def\comment#1{}

\begin{document}   
%\onecolumn[\hsize\textwidth\columnwidth\hsize\csname
%@onecolumnfalse\endcsname

\twocolumn[\hsize\textwidth\columnwidth\hsize\csname @twocolumnfalse\endcsname
\title{Small-Scale Cosmic Microwave Background Polarization from Reionization}
\author{Daniel Baumann$^{1,2}$, Asantha Cooray$^{1}$, and Marc
Kamionkowski$^{1}$}
\address{$^1$California Institute of Technology, Mail Code
130-33, Pasadena, CA 91125\\
$^2$University of Cambridge, Cambridge, CB3 OHE\\
E-mail: (db275,asante,kamion)@tapir.caltech.edu}
%\date{To be submitted to Phys. Rev. D. --- October 2001}

\maketitle

\begin{abstract}
We discuss fluctuations in the cosmic microwave background (CMB)
polarization due to scattering from reionized gas at low
redshifts.  Polarization is produced by re-scattering of
the primordial temperature anisotropy quadrupole and of the
kinematic quadrupole that arises from gas motion transverse
to the line of sight.  We show that both effects produce equal
E- and B-parity polarization, and are, in general, several
orders of magnitude below the dominant polarization
contributions at the last scattering surface to E-modes or the
gravitational-lensing contribution to B-modes at intermediate
redshifts.  These effects are also several orders of magnitude
below the B polarization due to lensing even after
subtraction with higher-order correlations, and are thus
too small to constitute a background for searches for the
polarization signature of inflationary gravitational waves.
\end{abstract}
\hfill

%\pacs{98.80.Es,95.85.Nv,98.35.Ce,98.70.Vc}      
]

\section{Introduction}

The angular power spectrum of cosmic microwave background (CMB)
temperature fluctuations is now becoming a powerful cosmological
probe \cite{Kno95}, both due to our detailed understanding of 
physics during the recombination era \cite{PeeYu70} and progress
on the experimental front \cite{Miletal99}.  In addition to
the primary anisotropies generated at the last-scattering
surface, CMB photons are also affected by large-scale
structure at low redshifts. These latter contributions result
from scattering off free electrons in clusters or the reionized
IGM and from modifications due the evolving gravitational field
associated with the formation of structures \cite{Coo02}. 

The existence of such secondary signals has now become
evident with the recent detection of small-scale power in
excess of that from primary fluctuations \cite{Masetal02}.  The
simplest and most plausible explanation for this small-scale
power is the Sunyaev-Zel'dovich (SZ; \cite{SunZel80}),
re-scattering of CMB photons from hot electrons in
unresolved galaxy clusters.  Although the
power observed is considerably larger than theoretical
expectations, the excess can be accommodated with a relatively
small increase in the power-spectrum amplitude
\cite{Bonetal,KomSel,CooMel,Sigetal02}.

In addition to small-scale anisotropies in the temperature,
increasing attention is now being devoted to detection of the
CMB polarization.  Besides resolving cosmological parameter
degeneracies \cite{Kno95}, the polarization will allow
several unique cosmological and astrophysical studies to be
carried out. These include a reionization signal
\cite{Zaldarriaga:1996ke}, probes of gravitational lensing \cite{SeljakZaldarriaga},
and a signature of inflationary gravitational
waves (IGW) through its contribution to the B, or curl, modes of
the polarization \cite{EB}.

Given the rapid pace of experimental progress and the rule of
thumb that the polarization is typically 10\% of the temperature
anisotropy, it is appropriate to investigate the polarization
produced by the secondary effects that have produced the
recently detected small-scale power.  This polarization is
produced by Thomson scattering of the quadrupole moment of the
radiation incident on the scatterer.  The quadrupole moment can
be either the primordial quadrupole that the scatterer sees
\cite{KamLoeb} or the kinematic quadrupole that arises from
quadratic terms in the Doppler shift when the gas moves
transverse to the line of sight \cite{SunZel80}.  Small-scale
angular fluctuations in this polarization are produced by
variations in the optical depth as a function of position across
the sky.

Since the secondary polarization signals can affect cosmological studies
involving the primordial polarization, and, by themselves, may
provide important information on astrophysics at late times, it
is important that we both quantify and understand the extent to
which large-scale structure can be a potential source of CMB
polarization.  Here, we discuss the angular power spectra for
scattering from reionized electrons and
study how these may affect potentially interesting studies with
CMB polarization. 

Our calculation parallels that of Ref. \cite{Hu99}, although
differs in that we present a simplified derivation of the
results based on a flat-sky approximation and use a
halo-clustering approach \cite{CooShe02} to describe
fluctuations in the electron density, in analogy to similar
recent calculations of
small-scale temperature fluctuations from unresolved clusters
\cite{KomSel,KomKit,Coo2000}.  We also include for the first
time the frequency dependence in the small-scale polarization
using results for the polarization from individual clusters
\cite{SazSun99,Challinor}.

The paper is organized as follows.  In \S~\ref{sec:pol}, we
introduce polarization signals associated with the
scattering of the primordial CMB temperature quadrupole and with
the kinematic quadrupole generated by transverse motions. We
then display and discuss our results in \S~\ref{sec:results}.
Though we present a general discussion
of the polarization, when illustrating results, we
will use the currently favored $\Lambda$CDM cosmology with
matter density (in units of the critical density)
$\Omega_m=0.35$, baryon density $\Omega_b=0.05$, vacuum-energy
density $\Omega_\Lambda=0.65$, Hubble constant (in units of 100
km~sec$^{-1}$~Mpc$^{-1}$) $h=0.65$, and spectral $n=1$ for
primordial density perturbations. We employ natural units
with the speed of light $c=1$ throughout.

\section{Polarization Power Spectra}\label{sec:pol}

If the radiation incident on a reionized electron has a nonzero
quadrupole moment, then the scattered radiation will be linearly
polarized.  The two dominant origins for this quadrupole
moment are: (a) A quadrupole from primordial CMB fluctuations, and
(b) a quadrupole from the quadratic term in the Doppler shift if
the scattering gas has a transverse velocity.

The statistics of the polarization produced in this way
is inherently non-Gaussian:  Although the amplitude of the
polarization in a given direction is determined by the optical
depth to Thomson scattering, the orientation is determined by
the radiation quadrupole incident on the scattering gas.
The power spectrum is thus in principle determined by a
convolution of the optical-depth field with the quadrupole
field.

In practice, however, we simplify by assuming that the
quadrupole moment is smooth over large distances, and that there
are small-scale variations only in the free-electron density.
We can then proceed to calculate the polarization fluctuations
that are linear in the electron-density fluctuations.  The
correlation length of the CMB quadrupole is comparable to the
horizon, and so our assumptions are fully justified for this
case, as long as we restrict our attention to multipole moments
$l \gtrsim10$.  On the other hand, the peculiar velocity is correlated
only on much smaller scale, $\sim$60 Mpc in our fiducial
$\Lambda$CDM cosmology.  Our calculations of the polarization
power spectrum for the kinematic quadrupole will thus be
reliable only for scales smaller than this, or multipole moments
$l\gtrsim100$.  

We first discuss the power spectra.  A polarization map will
consist of a measurement of the Stokes parameters $Q(\bn)$ and
$U(\bn)$ as functions of position $\bn$ on some patch of the
sky.  We can construct Fourier components $Q({\bf l})$ and
$U({\bf l})$ by
\begin{equation}
     X({\bf l}) = \int d^2 \bn~e^{i {\bf l} \cdot \bn } X(\bn) \, ,
\end{equation}
where $X \equiv Q, U$, and ${\bf l}$ is a vector in the plane of a region of the sky
sufficiently small to be considered flat.  Since the Stokes
parameters $Q$ and $U$ depend on the choice of axes, we consider
the rotationally invariant combinations,
\begin{eqnarray}
     E({\bf l}) &=& \cos(2\phi_{\mathbf l}) Q({\bf l})+\sin(2\phi_{\mathbf l})
     U({\bf l}) \nonumber \\
     B({\bf l}) &=& \cos(2\phi_{\mathbf l}) U({\bf l})-\sin(2\phi_{\mathbf l})
     Q({\bf l}) \, ,
\end{eqnarray}
where $\phi_{\mathbf l}$ is the angle between ${\bf l}$ and the chosen x-axis 
in the plane of the sky.  We define the angular power spectra
$C_l^{EE}$ and $C_l^{BB}$ from
\begin{equation}
     \langle Y(\vecl) Y(\vecl') \rangle = (2\pi)^2
     \delta_D(\vecl+\vecl') C_l^{YY} \, ,
\end{equation}
where $Y\equiv E,B$, and the angle brackets denote an average
over all realizations of the density field.

As discussed above, we suppose that the radiation quadrupole
moment is smooth over the region of sky we are considering.  If
so, then
\begin{eqnarray}\label{equ:EE}
     \langle E (\vecl) E(\vecl') \rangle &=& \langle
     B(\vecl)B(\vecl') \rangle  \nonumber \\
     &=& \frac{1}{2}  
     \left( \langle Q(\vecl) Q(\vecl') \rangle + \langle
     U(\vecl)U(\vecl') \rangle \right) \, ,
\end{eqnarray}
while
\begin{equation}\label{equ:EB}
     \langle E(\vecl)B(\vecl') \rangle = \frac{1}{2} \left(
     \langle Q(\vecl)U(\vecl') \rangle  -  \langle
     Q(\vecl)U(\vecl') \rangle \right) = 0 \, ,
\end{equation}
where the latter equality is consistent with parity
conservation.  These results can be derived by noting that if
the quadrupole moment is constant, then the orientation of the
polarization is constant.  If so, we may choose our axes on the
sky so that $U=0$.  Then, $E(\vecl) \propto \sin(2 \phi_{\mathbf l})$, and
$B(\vecl) \propto \cos(2 \phi_{\mathbf l})$, but when averaged over
the orientation angle $\phi_{\mathbf l}$, we recover Eq. (\ref{equ:EE}).

We now proceed to calculate the power spectra induced by
reionization.  The polarization in direction $\bn$ due to
scattering from free electrons is an integral along the line of
sight \cite{Kosowsky:1994cy},
\begin{equation}
     Q(\hat n) - i U(\hat n) = \sqrt{3 \over 40 \pi}
     \int \, dr {d \tau(r\bn,r) \over dr} a_{22}(r)\, ,
\label{eqn:projection}
\end{equation}
where $r$ is the comoving distance, $(d\tau/dr)(r\bn,r) =
\sigma_T n_e(r\bn,r) a(r)$, $a(r)$ is the scale factor
at a comoving distance $r$, $n_e(r \bn,r)$ is the
free-electron density at direction $\bn$ at distance $r$, and
$\sigma_T$ is the Thomson cross section.  

In Eq.~(\ref{eqn:projection}), 
$a_{22}(r)$ is the radiation quadrupole moment at distance $r$.  More
precisely, $a_{22}(r)$ is the coefficient of the
spherical harmonic $Y_{22}(\theta,\phi)$ in a spherical-harmonic
expansion of the radiation intensity in a coordinate system in
which the line of sight is the ${\bf \hat z}$ direction.  Note
that we take $a_{22}(r)$ to be a function of distance only, and
not direction, consistent with our assumption that the
quadrupole is coherent over a large patch of the sky.  
Since we use Limber's approximation below, in which angular correlations are
induced only by spatial separations at the same distance, the
variation of $a_{22}(r)$ with distance can be included
consistently.

With the polarization written as a projection,
Eq.~(\ref{eqn:projection}), along the line of sight, the angular
power spectrum follows in the flat-sky limit from Limber's
equation \cite{Limber},
\begin{eqnarray}
     C_l^{EE} &=& C_l^{BB} \nonumber \\
     &=& \frac{3}{80 \pi} 
     \int_0^{z_{\rm rei}} dz \, \frac{d^2V}{d\Omega dz} |a_{22}(z)|^2\,
     P^{(t)}_{\tau\tau} \left(\frac{l}{d_A},z\right) \, ,
\label{eqn:cl}
\end{eqnarray}
where $P^{(t)}_{\tau\tau}$ is the power spectrum of $d\tau/dr$,
proportional to the power spectrum of the electron density
$n_e$, and the integral is taken up to the redshift $z_{\rm rei}$ of reionization using
the comoving differential volume given by $d^2V/d\Omega dz$.

We now assume that the free-electron density is distributed like
the mass in the Universe and model the mass distribution
following the halo approach to large-scale structure
\cite{CooShe02}.  We thus decompose the power spectrum into two
parts, one that describes contributions from single halos
(1-halo term) and a part that accounts for 
correlations between halos (2-halo
term)~\cite{KomSel,KomKit,Coo2000,ColeKaiser}:
\begin{equation}
     P^{(t)}_{\tau\tau}= P^{1h}_{\tau\tau}+P^{2h}_{\tau\tau} \, ,
\end{equation}
where
\begin{eqnarray}
     P^{1h}_{\tau\tau}(k,z) &=&\int dM \frac{dn(M,z)}{dM} \left|
     \tau_l(M,z) \right|^2 \, , \nonumber \\
     P^{2h}_{\tau\tau}(k,z) &=&P^{\rm lin}( k, z)      \nonumber \\
      & & \times\left[\int dM
     \frac{dn(M,z)}{dM} b(M,z) \tau_l(M,z) \right]^2  \, .
\label{eqn:p}
\end{eqnarray} 
Here, $(dn/dM)(M,z)$ is the mass function of halos
\cite{Press:1973iz} and $b(M,z)=1+\frac{\nu^2(M,z)-1}{\delta_c}$
is the halo bias \cite{Mo:1995db} with
$\nu(M,z)=\delta_c/\sigma(M,z)$ the peak-height threshold
with rms fluctuation within a top-hat filter at the virial
radius corresponding to mass $M$ given by $\sigma(M,z)$ and
threshold overdensity of spherical collapse given by $\delta_c$.
Useful fitting functions for these quantities are tabulated in
Ref.~\cite{Henry:2002wr}.  

We define the projected scattering optical depth as
$\tau(\theta) = \sigma_T \int n_e(y,\theta)\, dy$
where $y$ is the line-of-sight distance along each halo at
angular distance $\theta$
from the cluster center. The two-dimensional Fourier transform
of $\tau(\theta)$ is,
\begin{equation}
\tau_l = 2\pi\int_0^{\theta_{\rm vir}} \theta d\theta\;
\tau(\theta) J_0(l\theta) \, ,
\end{equation}
where $\theta_{\rm vir}$ corresponds to the virial radius of the halo.
For simplicity, we model the electron distribution within each halo using a $\beta$-profile 
and normalize it such that the gas mass fraction of each halo produces the global baryon 
fraction \cite{KomKit}.
Finally, $P^{\rm lin}(k)$ is the power spectrum of the linear
density field. We use the formulae of Ref.~\cite{Hu:1998tj} to
describe the transfer function and normalize the power spectrum
to match $\sigma_8=0.9$ consistent with COBE fluctuations at
large scales \cite{Bunn:1996da}.  We comment below on the
$\sigma_8$ dependence of our results.

\begin{figure}[t]
\centerline{\psfig{file=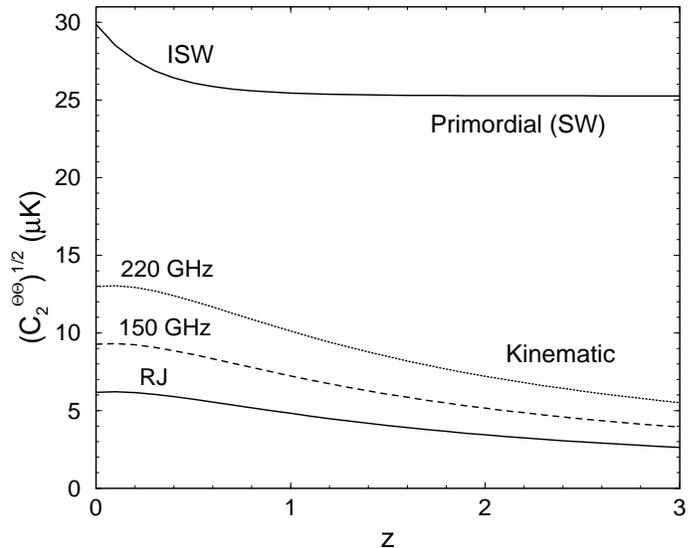,width=3.5in,angle=-90}}
\caption{The temperature quadrupole, $C_2^{\Theta \Theta}$, as a function of
     redshift. We show both the primordial
     and the kinematic quadrupole. The bottom
     kinematic-quadrupole curve is for $g(x)=1$, as appropriate
     for the Rayleigh-Jeans (RJ) part of the frequency spectrum, and
     the dashed and dotted curves are, respectively, for
     frequencies 150 and 220 GHz.}
\label{fig:PowSpec}
\end{figure}

\subsection{Primordial Quadrupole}

The primordial quadrupole will have a coherence length
comparable to the horizon.  We thus expect that the amplitude of
the polarization power spectrum measured on a ${\cal O}(10^\circ)$
patch of sky may differ by order unity from that on a different
${\cal O}(10^\circ)$ patch of sky (and likewise that our
calculation should break down for $l\lesssim10$).  However, when
averaged over the entire sky, the amplitude of the power spectrum should be
given by replacing the quantity $|a_{22}(z)|^2$ in
Eq.~(\ref{eqn:cl}) by its
expectation value, $C_2^{\Theta\Theta}(z)$, the variance of the temperature
quadrupole at redshift $z$.  At redshift $z=0$, with the
power-spectrum tilt fixed at unity,
COBE finds $C_2^{\Theta\Theta}(z=0)=(27.5 \pm 2.4 \,\mu{\rm K})^2$ 
\cite{Bennett:gg}. At
higher redshifts, the mean quadrupole moment evolves due to the
integrated Sachs-Wolfe effect and possibly, if the power
spectrum is not flat, due to any scale dependence since the
quadrupole probes smaller distances at earlier times.  We
calculate 
$C_2^{\Theta\Theta}(z)$ following Ref.~\cite{Hu99} for our fiducial
$\Lambda$CDM cosmology and show the result in Fig.~\ref{fig:PowSpec}.

Finally, note that Thomson scattering from
cold electrons will not change the photon frequency. Thus,
there will be no frequency dependence of the primordial quadrupole if we use
Stokes parameters in temperature units---rather than the
intensity units used by Ref.~\cite{SazSun99}---as we do
throughout this paper.

\begin{figure}[t]
\centerline{\psfig{file=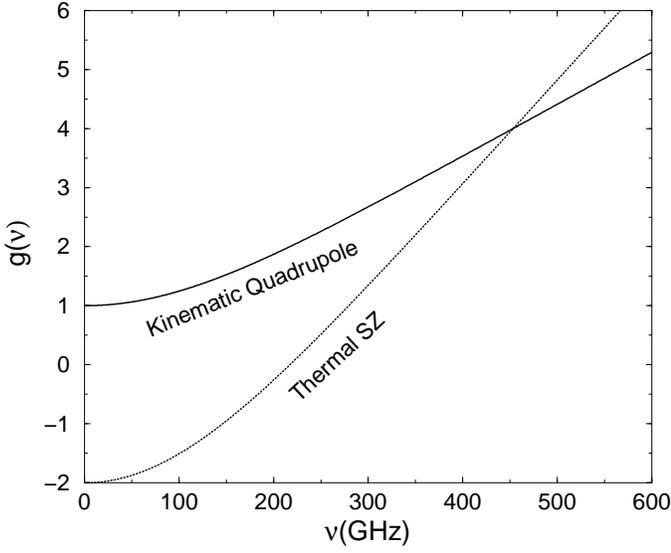,width=3.5in,angle=-90}}
\caption{The solid curve shows the frequency spectrum of the
     kinematic temperature (rather than intensity) quadrupole.
     For reference, we also show the spectral dependence of the
     SZ thermal effect with the dotted curve.}
\label{fig:freq}
\end{figure}

\subsection{Kinematic Quadrupole}

As discussed in Ref.~\cite{SunZel80}, an electron gas moving
with a transverse velocity $v_t$ relative to the CMB rest frame
sees an $a_{22}$ quadrupole moment (in a coordinate system in
which $z$ is along the line of sight),
\begin{equation}
     a_{22} = g(x) \sqrt{4 \pi \over 30} v_t^2\, e^{2 i \phi_v},
\end{equation}
where $\phi_v$ is the orientation angle for $v_t$ on the plane
of the sky.  To obtain this result, note that in a coordinate
system in which the ${\bf \hat z}$ axis is aligned with the
cluster's motion, the quadrupole dependence of the radiation
temperature is $g(x) v_t^2 (\mu^2-1/3) = g(x) v_t^2 (2/3)
\sqrt{4 \pi/5} Y_{2,0}$,
where $\mu$ is the cosine of the angle between the radiation
direction and the ${\bf \hat z}$ direction.  However, in a
coordinate system in which the ${\bf \hat z}$ axis is taken to be along the line
of sight, $(\mu^2-1/3)=-(1/3)\sqrt{4 \pi/5} Y_{2,0}+ \sqrt{2 \pi /
15}(Y_{2,2}+Y_{2,-2})$.  Thus, the coefficient of $Y_{2,2}$, the
component of the quadrupole moment that gives rise to
polarization in our direction, is only a fraction $\sqrt{3/8}$
of the total quadrupole moment.

Unlike the primordial quadrupole, the kinematic quadrupole has a frequency
dependence which we denote by $g(x)$, where $x=h\nu / kT_{\rm CMB}$.
Following Refs.~\cite{SazSun99,Challinor}, this frequency dependence can
be calculated by expanding the spectral intensity distribution
of the CMB in the rest frame of electrons,  $I_\nu \propto
x^3/(e^{x\gamma(1+v\mu)}-1)$, in terms of
velocity $v$, with $\gamma=(1-v^2)^{-1/2}$ and $\mu$ the cosine
of the angle between velocity and incident photon direction.  We
then obtain the frequency dependence of the quadrupole term, in
temperature units instead of intensity units, to be
$g(x)=(x/2)\coth(x/2)$. In Fig.~\ref{fig:freq}, we show the
frequency dependence of the kinematic quadrupole,
which was neglected in Ref.~\cite{Hu99} \footnote{Note
that our Figure for the frequency dependence differs from
Fig.~4 of Ref.~\cite{SazSun99} due to our use of temperature
units and their use of intensity units.}.
For reference, we also plot the frequency dependence of the
thermal Sunyaev-Zel'dovich effect,

The bulk flows associated with large-scale structure have coherence scales of
order $\sim$ 60 Mpc. At first, we might be tempted to think that
if we were to survey some region of sky small
compared with the coherence length $\sim1^\circ$ for the peculiar
velocity when projected at a typical redshift of order unity, 
then the power-spectrum amplitude might generally
differ by order unity from the power-spectrum amplitude on some
other similarly sized patch of sky.  This would be true if all
the fluctuations were produced at one distance.  However, since
there will be ${\cal O}(100)$ 60-Mpc coherence patches along
the line of sight, these will tend to average out, and the
polarization power-spectrum amplitude should be roughly the same
from one $1^\circ$ degree patch to another.  It is thus
appropriate to replace $|a_{22}(z)|^2$ in Eq.~(\ref{eqn:cl}) by
its expectation value,
\begin{equation}
     \VEV{|a_{22}|^2} = {4 \pi \over 30} g^2(x)\VEV{v_t^4} = {16 \pi
     \over 135} g^2(x) v_{\rm rms}^4\, ,
\label{eqn:vkin}
\end{equation}
where we have used
$\VEV{v_t^4}=\VEV{(v_x^2+v_y^2)^2}=(8/9) v_{\rm rms}^4$, since
$\VEV{v_x^2}=\VEV{v_y^2}=v_{\rm rms}^2/3$,
$\VEV{v_x^4}=\VEV{v_y^4}=3\VEV{v_x^2}^2=v_{\rm rms}^2/3$, and
$\VEV{v_x^2 v_y^2}=\VEV{v_x^2} \VEV{v_y^2}=v_{\rm
rms}^4/9$.  We calculate the linear-theory rms peculiar velocity
from
\begin{equation}
     v_{\rm rms}^2(r) = \int \frac{k^2 dk}{2 \pi^2} \left(
     \frac{\dot{G}}{k}\right)^2 P^{\rm lin}(k,0) \, ,
\end{equation}
where $G$ is the linear-theory density-perturbation growth
factor, and the overdot represents a derivative with respect to
radial distance.  According the halo-clustering
model \cite{CooShe02}, peculiar-velocity fields are correlated
over large distances and the nonlinear corrections to $v_{\rm
rms}^2$ are small. The resulting linear-theory rms kinematic
quadrupole is also shown in Fig.~\ref{fig:PowSpec} assuming
$g(x)=1$ as relevant for Rayleigh-Jeans (RJ) part of the
frequency spectrum when $x \rightarrow 0$, and for $\nu=150$ GHz
and $\nu=220$ GHz.

\begin{figure}[t]
\centerline{\psfig{file=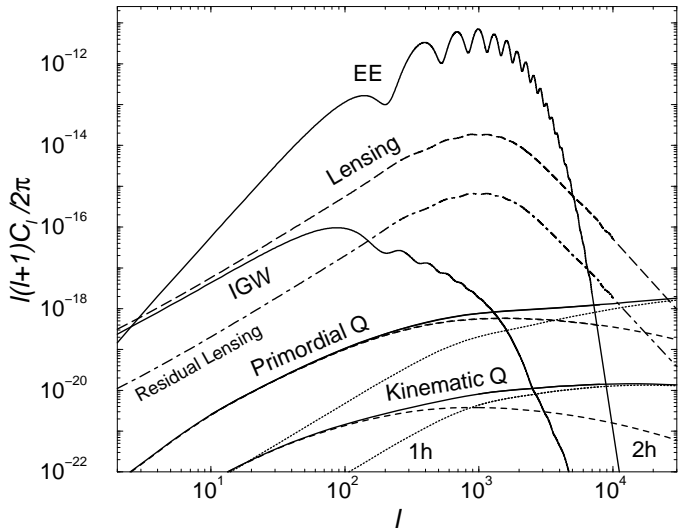,width=3.5in,angle=-90}}
\caption{
     Polarization power spectra due to the
     rescattering of the primordial and kinematic quadrupoles.
     We break total power in
     each of these cases into the 1- (1h; dotted lines) and 2-halo (2h; dashed lines) terms
     under the halo-based approach used here.  While at large
     angular scales correlations
     between halos dominate, at small angular scales of order few
     arcminutes and below, contributions are dominated by the
     1-halo term.  For comparison, we also show primordial
     polarization power spectra for E and B-modes involving
     dominant scalar (E-mode) and tensor (B-mode) contributions
     respectively. The tensor contribution to B-modes due to
     inflationary gravitational waves (IGW) assumes an energy scale
     for inflation of $10^{16}$ GeV. The long-dashed curve is the
contribution to B-modes of polarization resulting from the cosmic shear conversion of
power in E-modes, while 
the dot-dashed line labeled ``Residual Lensing'' represents the noise
contribution after optimally 
subtracting the lensing contribution using 
higher order statistics (see text for details).}
\label{fig:cl}
\end{figure}

Before moving on to discuss our results, a few words on how our derivations
compare with those of Hu~\cite{Hu99} are necessary.  Note that
our Eq.~(\ref{eqn:cl}) 
agrees with Eq.~(63) of Ref.~\cite{Hu99} after converting to the
logarithmic power spectrum, $\Delta^2(k) = k^3P(k)/2\pi^2$,
using Limber's approximation with $k=l/d_A$,  
and substituting $|a_{22}(z)|^2 = (4 \pi/5) Q^2_{\rm rms}$
for the temperature quadrupole. Similarly, our derivation for the
polarization power 
spectrum due to the kinematic quadrupole agrees with that of
Ref.~\cite{Hu99}, though 
this comparison requires an additional step. 
The kinematic quadrupole which replaces $Q^2_{\rm rms}$ in
Eq.~(63) of Ref.~\cite{Hu99} is given in
Eq.~(58) as $(8/45)(1-f_{\rm kin})v^4_{\rm rms}$. 
The factor $f_{\rm kin}$ accounts for mode coupling associated
with the velocity field
and is given by the mode-coupling integral in the second line of
Eq.~(58) of Ref.~\cite{Hu99}.
If we take the limiting case of $f_{\rm kin}$ with $y_1 \ll 1$
and $y_2 \rightarrow 1$,
such that the two velocity power spectra decouple from each
other, then $f_{\rm kin}=1/6$
exactly and the relevant kinematic quadrupole becomes $(8/54)
v^4_{\rm rms}$. This is consistent with Eq.~(\ref{eqn:vkin}) of
this paper.  This can be verified by substituting our
Eq.~(\ref{eqn:vkin}) into our Eq.~(\ref{eqn:cl}), which recovers
Eq.~(64) of Ref.~\cite{Hu99}.  The fact that Hu finds
numerically that $f_{\rm kin}=1/6$ to a very good approximation
justifies our assumption that the velocity and density fields
are effectively decoupled.

\section{Results and Discussion}
\label{sec:results}

\begin{figure}[t]
\centerline{\psfig{file=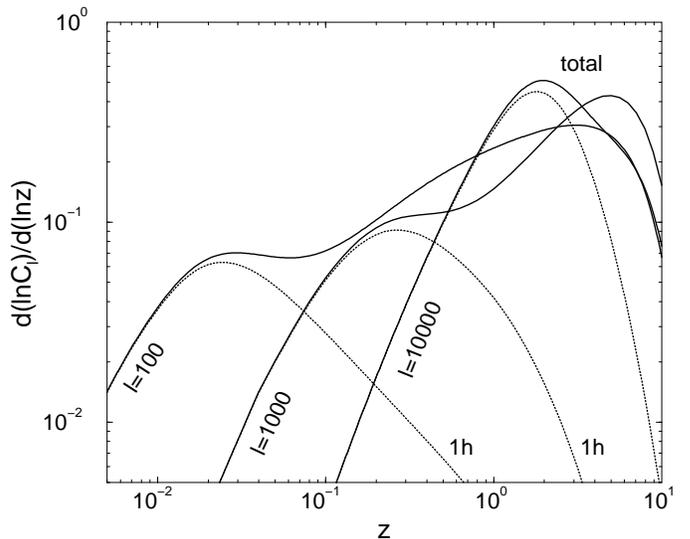,width=3.5in,angle=-90}}
\caption{The fractional contribution to polarization power
     spectra due to scattering of the primary anisotropy
     temperature quadrupole as a function of redshift.  Here, we
     plot $d \ln C_l/ d \ln z$, for three specific values of $l$
     (10$^2$, 10$^3$, and 10$^4$). We show the total (solid curves) as well as
      the 1-halo term (dotted curves). Note that
     contributions come from a broad range in redshift, while,
     with  increasing  $l$, or decreasing angular scale, fractional
contributions in the 1-halo term increase to higher redshifts.}
\label{fig:dCdz}
\end{figure}

\begin{figure}[t]
\centerline{\psfig{file=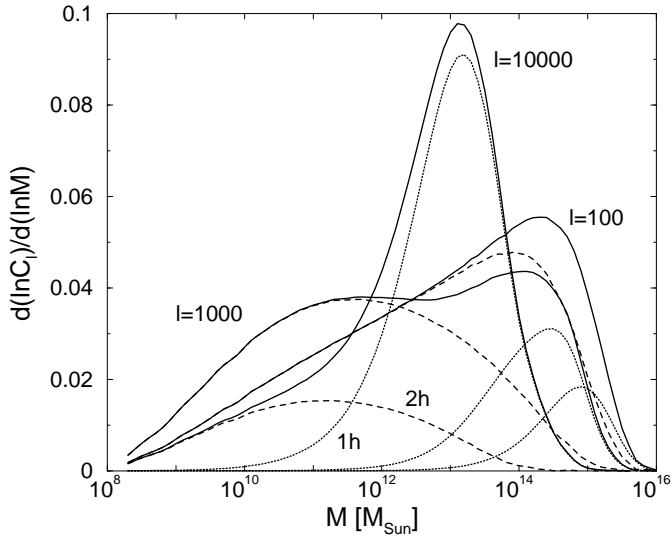,width=3.5in,angle=-90}}
\caption{The fractional contribution to polarization power spectra due to scattering of the
primary anisotropy temperature quadrupole as a function of cluster mass (in terms of solar mass).
Here, we plot $d \ln C_l/ d \ln M$, for three specific values of $l$ (10$^2$, 10$^3$, and 10$^4$). 
We show the total (solid line), the 1-halo term (dotted), and the 2-halo 
term (dashed).}
\label{fig:dCdM}
\end{figure}

We summarize our results on the polarization power spectra in
Fig.~\ref{fig:cl}.
Note that the secondary polarization discussed here contributes
equally to E- and
B-modes.  While the 1-halo term dominates at arcminute angular
scales and below, correlations between halos are important
and determine the total effect due to secondary
polarization at angular scales corresponding to a few
degrees.   The dependence of our results on the inclusion of the 2-halo
term is consistent with the result obtained for the temperature
power spectra from the kinetic SZ effect
\cite{Cooray:2001wa}, while it is inconsistent with the thermal
SZ effect, where contributions are dominated by the 1-halo term
over the whole range of angular scales. The latter behavior is
explained by the fact that the thermal-SZ effect is highly dependent
on the most massive halos, while the kinetic SZ effect, and the
secondary polarization signals calculated here, are {\it
in}dependent of the gas temperature and thus can depend on
halos with a wider mass range.

As shown in Fig.~\ref{fig:cl}, the secondary E-mode polarization
is several orders of magnitude below the E polarization from the
surface of last
scattering.  The secondary polarization is therefore
unlikely to be a source of confusion when interpreting
polarization contributions to E modes.  The amplitude of the
primary effect in B modes, due to gravitational waves, is highly
uncertain and depends on the energy scale of inflation \cite{KamKos}. For
illustration, we show in Fig.~\ref{fig:cl} the inflationary
gravitational wave (IGW) signal assuming an energy scale for
inflation of $E_{\rm infl}=10^{16}$ GeV; the amplitude of the
power spectrum scales as $E_{\rm infl}^4$.
At large angular scales the secondary polarization is several
orders of magnitude below the peak of this hypothetical IGW
polarization signal.  If the energy scale of inflation is
lowered considerably, say to $E_{\rm infl} \lesssim 10^{15}$
GeV, then we might guess that the secondary polarization could
ultimately constitute a background.

As also shown in
Fig. \ref{fig:cl}, however, there is a contribution to the B-mode power
spectrum that arises from conversion of the primary E modes to B
modes by gravitational lensing \cite{SeljakZaldarriaga}, and
this is considerably larger than the secondary polarization.
Moreover, we also show (the dot-dash curve) the contribution to
the irreducible B-mode power spectrum that remains even after
the lensing has been optimally subtracted with higher-order
correlations \cite{Kesdenetal,Knox,CooKes,Hu:2001fa}.  This residual lensing
power spectrum is considerably larger than the polarization from
reionization.  If the power spectrum is measured at a frequency
$\nu\simeq220$ GHz, then the polarization power spectrum from
the kinematic effect will be boosted by a factor $\sim5$ from
the $g(x)=1$ power spectrum shown in Fig. \ref{fig:cl}.
Moreover, if $\sigma_8=1$ (rather than the value $\sigma_8=0.9$
assumed in Fig.~\ref{fig:cl}), then both the secondary power
spectra will be increased for the same reasons that the
temperature power spectra increase by a factor
$\sim3$.\footnote{However, the boost in the polarization power
spectra will be smaller due to the fact that much of the
polarization is induced by electrons in smaller halos, rather
than the massive clusters that induce the temperature
fluctuation.}  Even with the possible frequency and $\sigma_8$
boosts, the secondary effects we consider here will be
unlikely to be a factor for either gravitational-lensing or
gravitational-wave studies with B modes.

In Fig.~\ref{fig:dCdz}, we show the fractional contributions to
polarization power spectra associated with the scattering of the
temperature quadrupole as a function of redshift.  We
show the total and the 1-halo term for three specific
values of $l$ corresponding to degree scales to arcminute
angular scales.  Though not shown here, we find consistent
behavior for the polarization power spectrum generated by the
scattering of the kinematic quadrupole.  As shown, contributions
come over a broad range in redshift out to the assumed
reionization redshift of 10 with a decrease 
at highest redshifts due to the decreasing
abundance of massive halos at higher redshifts.
At arcminute scales with $l \sim 10^4$, the signal
arises from halos at $z >1$. A comparison of
this behavior to Fig.~\ref{fig:PowSpec} reveals that at
redshifts greater than 1, the primary temperature quadrupole at the cluster positions results from a projection of the SW effect
only. Thus, at small angular scales, scattering contributions
come only from the quadrupole associated with the SW effect and
not the total that includes the ISW effect as well.

In Fig.~\ref{fig:dCdM}, we show the mass dependence of the
secondary polarization signal, again using the scattering
of the temperature quadrupole for illustration purposes. We show the
total, the 1-halo, and the 2-halo term at three different values
of $l$. While the 1-halo term is dominated by halos at the high-mass end of the mass function, the 2-halo term arises
from a wide range in halo mass. At tens of arcminute
scales equal contributions come from halo masses
in the range of 10$^{10}$ to 10$^{14}$ $M_{\sun}$. This is
consistent with the equivalent result for the kinetic SZ effect where
a wide range of masses contribute.

Note that in addition to the auto-correlation of polarization,
one expects secondary temperature fluctuations due to galaxy clusters
to be correlated with that of the secondary polarization involving the E-mode.
The temperature-polarization cross-correlation with the B-modes
is expected to be zero based on parity considerations.  We
considered all combinations between secondary polarization and
temperature anisotropies involving thermal and kinetic SZ
effects and found them to be zero based on simple geometric
arguments.  

While our simple flat-sky derivation of the secondary
polarization anisotropy from reionization agrees with
the all-sky approach
of Ref.~\cite{Hu99}, our calculational method complements the one used there.
We use the halo model to describe the non-linear power spectrum
of electrons while in Ref.~\cite{Hu99}, electrons of the
intergalactic medium were assumed to trace the
dark-matter-density field. Our numerical results agree well with
those of Ref.~\cite{Hu99}, particularly at large
angular scales where they should both converge to the same
linear-theory calculation.  Here we have neglected to consider
the expected smoothing of the electron density on small scales
from reheating of the IGM gas.  However, as Ref. \cite{Hu99}
shows, these effects are easily included and reduce the power
spectrum substantially only on angular scales $l \gtrsim10^4$
smaller than those we have considered here.

While our halo-based approach is likely to be affected by uncertainties
related to the mass function or the distribution of electrons within
halos, we expect our calculations to accurately reflect the polarization
anisotropy power at small angular scales. In the case of the kinematic quadrupole,
it is likely that we have overestimated the power at scales of a few degrees or more
due to our assumption that the velocity field is coherent at such angular scales.
We expect this assumption to only affect the 2-halo term of correlations
and to result in an overestimate of $C_l$ when $l$ is less than $\sim$ 1000.

\acknowledgments

We thank K.~Sigurdson for useful discussions.
This work was supported in part by NSF AST-0096023, NASA
NAG5-8506, and DoE DE-FG03-92-ER40701.

\end{document}